\documentclass[useAMS,usenatbib]{mn2e}

\usepackage{verbatim}
\usepackage[english]{babel}
\usepackage{graphicx}
\usepackage{color}
\usepackage{amssymb}
\usepackage{amsfonts}
\usepackage{amsmath}

\DeclareOldFontCommand{\sc}{\normalfont\scshape}{\@nomath\sc}
\newcommand{\h}{H}
\newcommand{\hi}{H\,{\sc i}}

\newcommand{\he}{He}
\newcommand{\hei}{He\,{\sc i}}
\newcommand{\heii}{He\,{\sc ii}}
\newcommand{\heiii}{He\,{\sc iii}}

\newcommand{\mh}{$h^{-1}\textrm{Mpc}$}
\newcommand{\kh}{$h^{-1}\textrm{kpc}$}

\newcommand{\kms}{$\tr{km}\,\tr{s}^{-1}$}

\newcommand{\tr}{\textrm}
\newcommand{\ti}{\textit}
\newcommand{\be}{\begin{equation}}
\newcommand{\ee}{\end{equation}}
\newcommand{\bea}{\begin{eqnarray}}
\newcommand{\eea}{\end{eqnarray}}

\newcommand{\aj}{AJ}
\newcommand{\apj}{ApJ}
\newcommand{\apjl}{ApJL}
\newcommand{\apjs}{ApJS}
\newcommand{\mnras}{MNRAS}
\newcommand{\aap}{A\&A}
\newcommand{\araa}{ARA\&A}
\newcommand{\nat}{Nature}
\newcommand{\pasa}{PASA}

\title[AGN-driven helium reionization]{AGN-driven helium reionization and the incidence of extended \heiii\ regions at redshift $z>3$}

\author[M. Compostella, S.Cantalupo, \& C. Porciani]{Michele Compostella$^{1}$\thanks{E-mail:\newline
mcompos@astro.uni-bonn.de (MC);\newline cantal@ucolick.org (SC); \newline porciani@astro.uni-bonn.de (CP).}, Sebastiano Cantalupo$^{2}$\footnotemark[1] and Cristiano Porciani$^{1}$\footnotemark[1]
\\
$^{1}$Argelander Institut f\"{u}r Astronomie der Universit\"{a}t Bonn, Auf dem H\"{u}gel 71, Bonn, D-53121, DE.
\\
$^{2}$Department of Astronomy and Astrophysics, UCO/Lick Observatory, University of California, 1156 High
Street,\\ Santa Cruz, CA 95064, USA.
}

\begin{document}

\date{Accepted ; Received ; in original form .}

\pagerange{\pageref{firstpage}--\pageref{lastpage}} \pubyear{2014}

\maketitle

\label{firstpage}

\begin{abstract}
We use hydrodynamic simulations post-processed with the radiative-transfer code {\sevensize RADAMESH} to assess recent claims that the low \heii\ opacity observed in $z > 3$ quasar spectra may be incompatible with models of \heii\ reionization driven by the observed population of active galactic nuclei (AGNs). In particular, building upon our previous work, we consider an early population of sources and start the radiative-transfer calculation at redshifts $z\geq 5$. Our model faithfully reproduces the emissivity of optically selected AGNs as inferred from measurements of their luminosity function. We find that \heii\ reionization is very extended in redshift ($\Delta z\geq 2$) and highly spatially inhomogeneous. In fact, mock spectra extracted from the simulations show a large variability in the evolution of the \heii\ effective optical depth within chunks of size $\Delta z=0.04$. Regions with low opacity ($\tau_\tr{\heii}^\tr{eff}<3$) can be found at high redshift, in agreement with the most recent observations of UV-transmitting quasars. At the highest redshift currently probed by observations ($z\sim 3.4$), our updated model predicts a much lower \heii\ effective optical depth than previous simulations in the literature relieving most of the tension with the current data, that, however, still persists at about the (Gaussian) $1\sigma$ to $2\sigma$ level. Given the very small number of observed lines of sight, our analysis indicates that current data cannot rule out a purely AGN-driven scenario with high statistical significance.
\end{abstract}


\section[]{INTRODUCTION}
The first few Gyr of cosmic history witnessed at least two major changes in the main constituents of the intergalactic medium (IGM). The ionized fraction of hydrogen and helium dramatically increased in a relatively short time-scale passing from nearly fully neutral to fully ionized \citep[see][for a detailed review]{Fan2006a, Meiksin2009}. The occurrence and timing of these events - called cosmic reionizations - are connected with the density, luminosity and spectral hardness of the first photoionizing sources, e.g. stars in galaxies and active galactic nuclei (AGNs). Uncovering when and how these processes took place can therefore provide key information for both the study of the high-redshift IGM and the luminous component of the Universe.

The reionization of \hi\ (and \hei) was completed by redshift 6, as evidenced by the lack of Gunn-Peterson absorption troughs in the spectra of high-redshift quasars \citep[e.g.][]{Fan2006b}. It is currently believed that cosmic reionization of \heii\ is delayed to later epochs.
This expectation is based on the following arguments: (i) normal star-forming galaxies produce a large amount of photons with energies above 1 ryd - thus ionizing \hi\ - but do not contribute substantially to the cosmic ionizing background above 4 ryd, the ionization threshold of \heii; (ii) among all sources with harder spectra, AGNs emit enough radiation to bring \heii\ reionization to completion but their abundance declines much more rapidly than the spatial density of star-forming galaxies at $z>4$.

Similarly to \hi\ reionization studies, the main observational probes of the final stages of the \he\ reionization process come from the study of the \heii\ Ly$\alpha$ forest in the spectra of high-redshift quasars. Unfortunately, the paucity of the sources for which the far ultraviolet (UV) is not extinguished by intervening \hi\ absorption and the need to observe from space (given the \heii\ Ly$\alpha$ rest-frame wavelength of $304\,\tr{\AA}$) resulted to date in a handful of useful lines of sight for \heii\ reionization studies at $2.5<z<3.5$ \citep[][]{Syphers2009,Shull2010,Worseckand2011,Worseck2011,Syphers2012}.
Since the discovery of the first `\heii\ quasar' Q0302-003, observations with the \ti{Far Ultraviolet Spectroscopic Explorer} (\ti{FUSE}) and the \ti{Cosmic Origins Spectrograph} (\ti{COS}) on the \ti{Hubble Space Telescope} (\ti{HST}) revealed that the mean \heii\ opacity at $z\sim 3$ is much higher than the opacity of \hi\ at similar redshifts \citep[][]{Jakobsen1994,Syphers2009,Worseck2011} with a slow recovery of transmitted flux at $z<2.7$. While this suggests that \heii\ reionization was indeed completed later than \hi\ reionization, the large scatter and variance between effective \heii\ opacities measured along different sightlines in a wide redshift range indicate that \heii\ reionization may have been a very extended process in time \citep{Syphers2011a,Syphers2011b,Worseck2011,Syphers2014}. The presence of sightlines showing moderate opacity and some transmitted flux at $z\gtrsim 3$ is of particular interest to understand when \heii\ reionization was still ongoing. For $3<z<3.5$, typical values for the effective \heii\ optical depth are around $5$ \citep{Heap2000,Zheng2004,Syphers2011a,Worseck2011,Syphers2014}. Moreover, very recent \ti{COS} observations of 8 new sightlines at $z\sim 3$ and the re-analysis of 11 archival \ti{COS} spectra by \citet{Worseck2014} suggest that the large line-to-line variation and low effective opacities (with optical depths as small as $2$) may persist to the highest redshift probed to date, i.e. $z\sim 3.5$. However, these measurements are difficult and require sizable corrections which make them prone to systematic errors. For this reason, no consensus has been reached yet on the transparency of the \heii\ forest at $z>3.2$ (Syphers, private communication).

The presence of transmitted regions in \heii\ spectra at such a high redshift questions the validity of current models in which \heii\ reionization is triggered by AGN activity only \citep{Worseck2014}. If the high transmissivity of the IGM at $z\sim 3.5$ is confirmed by future studies, it might be necessary to consider the contributions of additional, more exotic, sources in order to reconcile theoretical model and observations. For instance, normal galaxies emit copious soft and hard X-ray photons produced by X-ray binaries \citep{Ranalli2003,Gilfanov2004,Mineo2012,Basu-Zych2013}. The bolometric luminosity of these sources is expected to be several orders of magnitude below the instantaneous luminosity of AGNs, but their time-integrated contribution may be non-negligible, at least in the X-ray band \citep[e.g.][]{Power2009,Power2013,Mirabel2011,Fragos2013}. However, very little is known about the number of ionizing photons contributed by these sources in the far-UV. Alternatively, shocks due to cosmological structure formation have also been proposed as major contributors to the cosmic UV background above $4$ ryd \citep[e.g.][]{Miniati2004}. If these sources are important, \heii\ reionization may have started with very little or no delay with respect to \hi\ reionization.
It is worth stressing, however, that, while the bright end of the AGN luminosity function is well constrained by current observations, very little is known about the evolution of the fainter AGNs at $z > 3.5$ \citep[][]{Fontanot2007,Glikman2011,Giallongo2012} and their contribution to the total UV background. Uncertainties in the spectra, opening angle, lifetime and luminosity function make the UV-emissivity of AGNs not well determined. Moreover, yet undetected episodes of black hole accretion that take place in small galaxies at very high redshift \citep[e.g.][]{Ricotti2004} are not accounted for in the most popular models.

In this paper, we assess the statistical significance with which the data by \citet{Worseck2014}, if corroborated by further evidence, would rule out the minimal AGN-based models of \heii\ reionization that have been presented in the literature. These models completely rely on the observed population of AGNs and make very little assumptions. We address this problem using radiative-transfer simulations of AGN-driven \heii\ reionization based on an updated version of the model presented in \citet[][hereafter Paper~I]{Compostella2013}.
In Paper~I, we used high-resolution adaptive mesh refinement (AMR) hydrodynamic simulations to investigate the topology of the \heii\ reionization produced by a population of AGNs that illuminate the computational volumes between $z=4$ and $z=2.5$.
The initial redshift for the radiative-transfer calculation was chosen to minimize the computational time and some numerical inaccuracies \citep[namely those due to neglecting the hydrodynamic response of the IGM to photoionization, which came out to be much smaller than we originally expected, see appendix A in Paper~I and][]{Meiksin2012} given that previous observations and models suggested that \heii\ reionization took place around $z\sim 3$.
We showed that the reionization process is highly inhomogeneous and extends for a redshift interval $\Delta z\gtrsim 1$. During this time, the IGM is composed of two phases: sharp ionization fronts expanding around AGNs separate hot regions of completely ionized gas from cold regions where \he\ is still singly ionized. The resulting patchiness of the reionization process gives rise to a bimodal distribution of the IGM temperature and produces two distinct equations of state for the singly ionized and the fully ionized medium.

Mock spectra of the \heii\ Ly$\alpha$ forest extracted from the simulations presented in Paper~I were in excellent agreement with low-redshift data. However, for the redshift interval $z>3.2$ in which data were not available at the time, our model was predicting a rapid increase in the median effective optical depth of small patches in the IGM \citep[see also][]{McQuinn2009}.
This fast change was driven by our choice of starting the radiative-transfer calculations at $z=4$. In fact, early AGNs, although rare, emit enough energetic photons to produce a substantial \heiii\ fraction by $z=4$ (see Fig.~\ref{fig:reion}).
In this work, we expand our previous analysis considering the same AGN population used in Paper~I but starting now our radiative-transfer computation at much earlier epochs, i.e. $z=5$ and $z=6$, with an updated version of our model. In particular, we focus our attention to the evolution of the \heii\ effective optical depth at high redshift and to the direct comparison with the new data that recently became available.

The paper is organized as follows. In Section 2, we describe the numerical simulations adopted in this work and discuss the main properties of the radiative-transfer calculations, focusing on the differences with Paper~I. We present the results extracted from mock spectra of the \heii\ Ly$\alpha$ forest in Section 3, where we compare the evolution of the effective optical depth with the most recent observational data at high redshift. In the same section, we investigate the properties of regions of transmitted flux in \heii\ spectra at high redshift. Finally, we summarize our findings and conclusions in Section 4.

\section[]{METHODS}
\label{sec:methods}

\subsection{Hydrodynamic simulations}
We perform a set of hydrodynamic simulations of the IGM with an upgraded version of the publicly available {\sevensize RAMSES} code \citep{Teyssier2002}. We adopt a flat $\Lambda$ cold dark matter cosmology with matter density parameter $\Omega_{\rm m}=0.2726$, baryon density parameter $\Omega_{\rm b}=0.0456$, and present-day value of the Hubble constant $H_0=100\,h~\tr{km}\,\tr{s}^{-1}\,\tr{Mpc}^{-1}$ with $h=0.704$. Primordial density perturbations are described by a Gaussian random field with a scale invariant spectrum characterized by the spectral index $n=0.963$ and a linear rms value within $8\,\tr{\mh}$ spheres of $\sigma_8=0.809$. Our assumptions are consistent with results from the \textit{Wilkinson Microwave Anisotropy Probe} presented in \citet{Jarosik2011}.

The simulated volume consists of a periodic cubic box $100\,\tr{\mh}$ on a side and is originally discretized into a regular Cartesian mesh with $256^3$ elements.
We adopt a quasi-Lagrangian AMR strategy based on local density: cells that contain more than eight dark-matter particles (or the proportionate baryonic mass) are tagged for refinement.
As a result, six further levels of refinement on top of the base grid are added at $z=4$ reaching an effective resolution of $16384^3$ in the densest regions (corresponding to a cell size of $6\,\tr{\kh}$).

We evolve the initial conditions from redshift $120$ to $4$ assuming that the gas has an ideal equation of state with adiabatic index $\gamma=5/3$ and a chemical composition produced by big bang nucleosynthesis with a helium mass fraction of $0.24$. We neglect small-scale phenomena like star formation, metal enrichment and supernova feedback which have limited impact on the reionization of the intergalactic \he.

We model soft UV photons emitted by star-forming regions in galaxies using a spatially uniform, but time-varying, background. We determine the photoionization and photoheating rates of this background by rescaling the model predictions by \citet{Haardt2012} with a multiplicative factor $C_\tr{g}(z)$ in order to match the opacity of the IGM measured in observational studies of the \hi\ Ly$\alpha$ forest (see Appendix \ref{app:uvbackground} for further details).

\subsection{Numerical radiative transfer}

\begin{table}
\centering
\begin{tabular}{l c c c c c}
\hline
Run & $L_\tr{box}$& $\Delta L_\tr{min}$ $^a$ & $\Delta
L_\tr{max}$ $^b$ & AGN & $z_\tr{AGN}$\\
&$\tr{(\mh)}$ &$\tr{(\kh)}$ & $\tr{(\kh)}$ & model & \\[0.5ex]
\hline
E1a & $100$  & $390$ & $6$  & PLE & $5$\\
E1b & $100$  & $390$ & $6$  & PLE & $5$\\
E1c & $100$  & $390$ & $6$  & PLE & $5$\\[1.ex]
E2a & $100$  & $390$ & $6$  & PDE & $5$\\
E2b & $100$  & $390$ & $6$  & PDE & $5$\\
E2c & $100$  & $390$ & $6$  & PDE & $5$\\ [1.ex]
P1a & $100$  & $390$ & $6$  & PLE & $6$\\
P2a & $100$  & $390$ & $6$  & PDE & $6$\\
\hline
\end{tabular}
\\
\begin{flushleft}
\textbf{Notes:} \\
$^a$ Size of a resolution element associated with the base grid.\\
$^b$ Size of a resolution element associated with the highest level of
refinement.\\
\end{flushleft}

\caption{Summary of the simulations used in this work.}
\label{table:sims}
\end{table}

We study the transmission of hard UV radiation through the IGM by post-processing our hydrodynamic simulations with a modified version of the {\sevensize RADAMESH} code \citep{Cantalupo2011}: photons emitted by a realistic population of AGNs are propagated through the computational volume allowing for periodic boundary conditions. We sample the spectrum of the sources using $50$ frequency bins ($10$ for \hi, $10$ for \hei\ and $30$ for \heii) logarithmically spaced in the energy range $1-40\,\tr{ryd}$ and we limit the propagation of the ionization fronts to the speed of light. In Paper~I, we have shown that neglecting the hydrodynamic response of the gas only generates minor artifacts that do not affect our results in any significant way \citep[see also][]{Meiksin2012}.

We use a probabilistic algorithm to populate the simulated dark-matter haloes at $z=4$ with AGNs (see Paper~I for further details). This procedure makes sure that their luminosity function accurately reproduces the measurement by \citet{Glikman2011}. The resulting total AGN emissivity at $z=4$ is roughly $25$ per cent higher than in \citet{Haardt2012}. We assume a simple `light-bulb' model, where an AGN shines for 45 Myr with a constant luminosity and then switches off. The spectral energy distribution of the emitted radiation is described by a broken power law with low- and high-frequency indices that are Monte Carlo sampled from the Gaussian distributions given in \citet{VandenBerk2001} and \citet{Telfer2002}, respectively.

We always use the $z=4$ snapshot of the hydrodynamic simulations to perform the radiative-transfer calculations since the AGN luminosity function by \citet{Glikman2011} has been measured at this redshift. We thus keep the baryonic overdensity $\Delta_{\rm b}(\mathbf{x})=\rho_{\rm b}(\mathbf{x})/\langle\rho_{\rm b}\rangle$ fixed during the radiative-transfer calculations and scale the mean density $\langle\rho_{\rm b}\rangle$ proportionally to $(1+z)^3$. Contrary to Paper I, we consider the contribution of an early population of AGNs which turn on at redshift $z_\tr{AGN}=5$. Therefore, a choice needs to be made regarding the initial gas temperature and the ionized fractions of \h\ and \he\ to input to {\sevensize RADAMESH}.
We determine them by enforcing photoionization equilibrium with the UV background at $z=5$
\footnote{The resulting temperature-density relation changes very little between redshifts 5 and 4. For instance, the temperature at mean density decreases by less than $500\,\tr{K}$.}.

To account for the evolution of the AGN emissivity with time we use the results from \citet{Haardt2012} (rescaled upwards by $25$ per cent to be consistent with the AGN luminosity function at $z=4$ measured by \citealt{Glikman2011}) and consider two different approximations. In the Pure Luminosity Evolution (PLE) model, the luminosity of the newly active sources that are associated with haloes of mass $M$ changes with time, while the total number of active sources in the computational volume scatters around a constant mean value.
Conversely, in the Pure Density Evolution (PDE) model, the luminosity of the sources does not evolve with time, while the average number of active sources grows monotonically in the redshift range considered in this work. Further details about both source models are given in Paper~I.

We perform three different simulations for each evolutionary model, for a total of six runs. To address the effect of an earlier start of the radiative-transfer calculations, we also perform one run for each source model with $z_\tr{AGN}=6$. In Table~\ref{table:sims}, we summarize the main characteristics of our simulations.

\section[]{RESULTS}
In the remainder of this paper, we present the results obtained averaging over all simulations of the E series. In fact, the PLE and PDE runs do not show any systematic difference in the average and scatter of the observables of interest.

\subsection{Reionization History}

\begin{figure}
\centering
\includegraphics[width=0.5\textwidth]{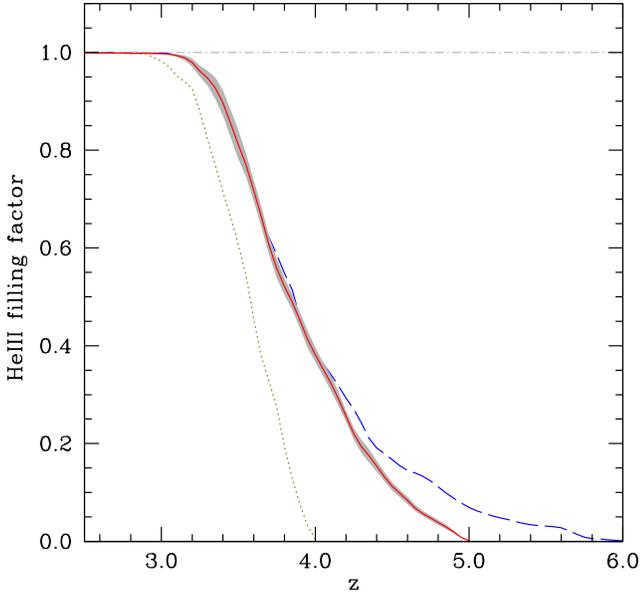}
\caption{Redshift evolution of the \heiii\ filling factor in our simulations. The red solid line represents the average in bins of $\delta z=0.05$ of the simulations with $z_\tr{AGN}=5$ presented in this work. The shaded region denotes the standard error of the mean. The blue dashed line indicates the evolution of the mean filling factor in the P simulations. The average of our results from Paper~I is also shown for comparison (green dotted line).}
\label{fig:reion}
\end{figure}

In Fig.~\ref{fig:reion}, we show the evolution of the \heiii\ filling factor ($f_\tr{\heiii}$, i.e. the volume-weighted average of the local \heiii\ fraction) in the E series. At $z=5$, when we start the radiative-transfer calculations, the abundances of \h, \he\ and their ions are set by the soft UV background generated by young stars. The initial \hi\ filling factor is $\sim 1\times 10^{-5}$ and most of the helium is singly ionized. Only $0.1$ per cent of the \he\ atoms are either neutral or completely ionized.
Turning on an AGN generates an ionization front that rapidly processes and heats the gas elements it crosses. The geometry of the resulting \heiii\ region is shaped by the local distribution of matter and reaches a comoving size of several \mh\ before the AGN switches off. At this point, the gas in the ionized bubble cools and the \heiii\ recombines until new ionization fronts reach the same spatial location.
Eventually, different ionization bubbles percolate and progressively fill the entire volume. When the reionization process is concluded, the mean temperature of the gas slowly decreases with time due to Hubble cooling.

\begin{figure}
\centering
\includegraphics[width=0.50\textwidth]{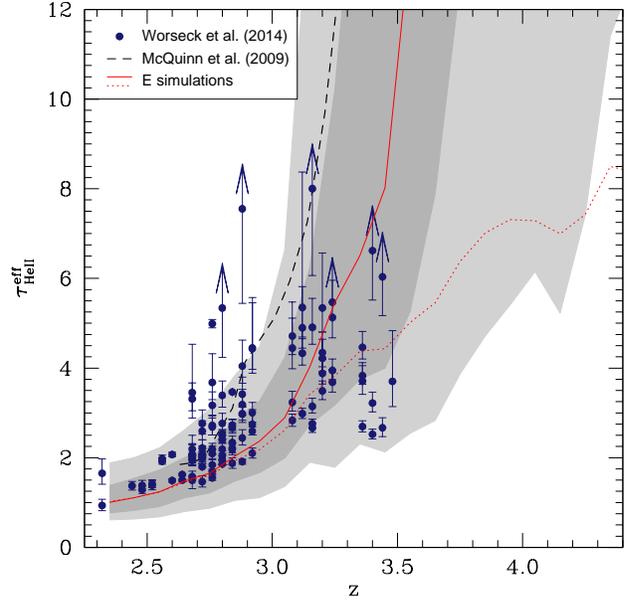}
\caption{Redshift evolution of the \heii\ effective optical depth in our simulations using bins of $\Delta z=0.04$. Shaded areas denote the regions of the $\tau_\tr{\heii}^\tr{eff}$ distribution enclosed between the $16$th and $84$th percentile (dark grey) and between the second and $98$th percentile (light grey). The red solid line corresponds to the median value and the red dotted line represents the mean effective optical depth averaged in bins of width $\delta z=0.1$ to reduce random fluctuations. Observational data by \citet{Worseck2014} are overplotted for comparison. The black dashed line indicates the median $\tau^\tr{eff}_\tr{\heii}$ extracted from the radiative-transfer simulation L3 by \citet{McQuinn2009}.}
\label{fig:tauHe}
\end{figure}

At $z=4$, $f_\tr{\heiii}\sim 0.38$ and similar values are found also in the simulations of the P series (Fig. \ref{fig:reion}). This suggests that \heii\ reionization is a process extended in time which requires an earlier start than assumed in Paper I (where we used $z_\tr{AGN}=4$). Note that the E and P series give very similar results for $z<4$ implying that our findings based on the E simulations should be robust in this redshift range.

The runs with $z_\tr{AGN}>4$ show a shallower evolution of $f_\tr{\heiii}$ at early times with respect to Paper I. This is due to the combined action of two
effects. First, the integrated quasar emissivity declines at high redshifts. Secondly, the reionization process is more inefficient in a denser medium because, in the pre-overlap era, individual \heiii\ regions quickly recombine when the central source switches off.

The end of the reionization process, defined as the first time \heiii\ accounts for $99.5$ per cent of the total \he\ content, is achieved at redshift $\sim 3$ and the \heiii\ filling factor increases above $99.9$ per cent only at $z<2.8$.
This is in good agreement with the results presented in Paper~I, where the late start of the radiative-transfer calculations causes only a slight delay in the completion of the reionization process.

\subsection{Effective optical depth}\label{sec:optical_depth}

\begin{figure}
\centering
\includegraphics[width=0.5\textwidth]{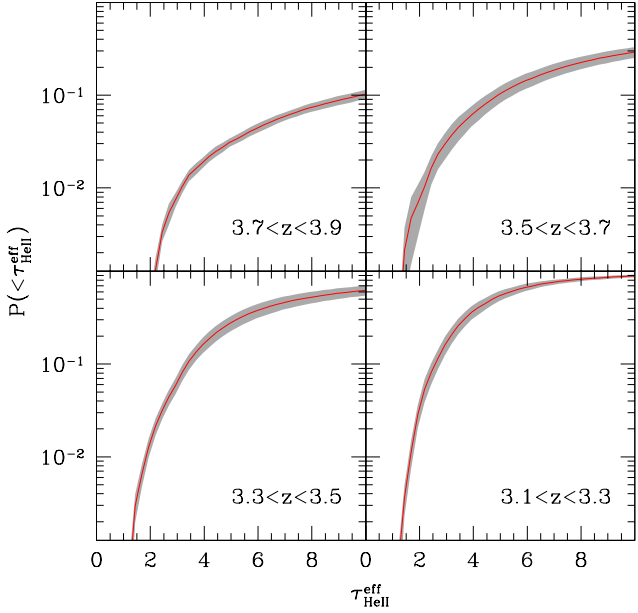}
\caption{Cumulative distribution of the \heii\ effective optical depth for different redshift intervals. Solid lines represent the average probability in bins of $\delta \tau^\tr{eff}_\tr{\heii}=0.25$ extracted from our simulations. Shaded regions correspond to the standard error of the mean between the different realizations.}
\label{fig:cumulative}
\end{figure}

\begin{figure}
\centering
\includegraphics[width=0.5\textwidth]{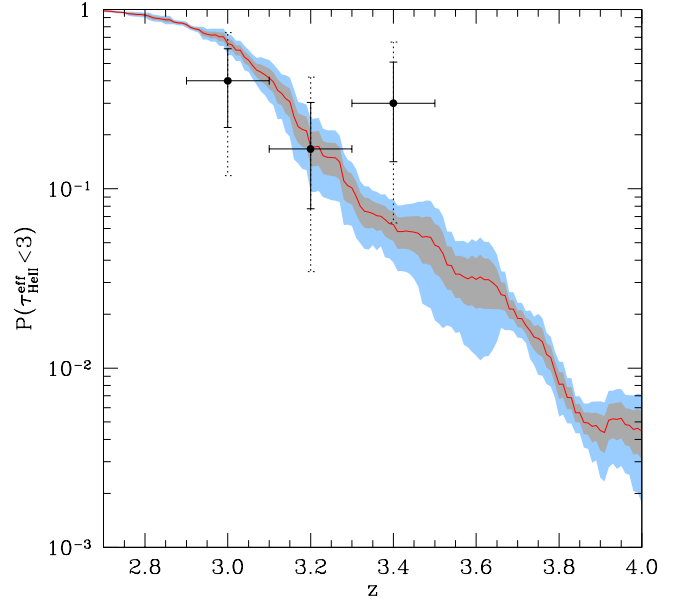}
\caption{Probability as a function of redshift of measuring $\tau^\tr{eff}_\tr{\heii}<3$ in our simulations. The solid line represents the running average of the different realizations in bins of $\delta z=0.2$. Shaded regions correspond to $1$ (grey) and $2$ (cyan) times the standard error of the mean. The probability extracted from observational data by \citet{Worseck2014} together with the two-sided $0.683$ (solid) and $0.955$ (dotted) confidence level for a binomial distribution are also reported for comparison. Note that the lines do not include the corrective factor for the observability bias. As discussed in Section \ref{sec:obsbias}, this bias will boost the simulated probability by about a factor 1.2 at $z=3.4$, bringing the simulations in better agreement with the observed fraction of transparent chunks.}
\label{fig:probtau}
\end{figure}

From each simulation, we extract synthetic Ly$\alpha$ absorption spectra considering the distribution and velocity of \heii\ along $100$ sightlines that are parallel to one of the main axes of the box and randomly located in the perpendicular plane. In order to mimic the instrumental response of a real spectrograph, we adopt a resolution element of $1\,\tr{\kms}$ and we convolve each spectrum with a Gaussian profile characterized by a kernel with a full width at half-maximum of $88\,\tr{\kms}$. This value corresponds to the nominal resolution of our simulations in the low density regions at $z=3.5$.
Note that our mock spectra are not able to resolve the single absorption features of the Ly$\alpha$ forest, but provide information on the coarse-grained properties of the medium and on the interaction between the sources and the surrounding gas.

We want to compare measurements of the IGM opacity extracted from our spectra against real data from observations of the \heii\ Ly$\alpha$ forest. For this reason, we evaluate the mean transmitted flux $\bar{F}_\tr{i}(z)$ ($0\leq \bar{F}_\tr{i}(z)\leq 1$) in spectral `chunks' of size $\Delta z=0.04$ (here the subscript $i$ runs over the different chunks, note that each line of sight includes several of them). We then compute the probability density function of the effective optical depth, $\tau^{\tr{eff}}_\tr{i}(z)=-\ln \bar{F}_\tr{i}(z)$, considering the chunks (from all simulations) that are contained within a redshift bin of width $\delta z=0.1$. Finally, we define the mean effective optical depth as $\bar{\tau}^{\tr{eff}}_\tr{\heii}(z)=-\ln \langle \bar{F}_\tr{i}(z)\rangle_\tr{i}$ where the mean is taken over the ensemble of chunks that lie within a redshift bin.

In Fig.~\ref{fig:tauHe}, we compare the distribution of the \heii\ effective optical depth derived from the E simulations with the observational data points recently presented in \citet{Worseck2014}. The shaded regions indicate the range of the simulated $\tau^{\tr{eff}}_\tr{i}(z)$ which is enclosed between the $16$th and the $84$th percentiles (dark grey) and between the second and the $98$th percentiles (light grey).
The red solid line shows the median effective optical depth as a function of redshift, while the red dotted line indicates the evolution of mean effective optical depth.
For comparison with previous work, we also report the evolution of the median $\tau^{\tr{eff}}_\tr{\heii}(z)$ as predicted by the simulations described in \citet[][]{McQuinn2009}, reanalyzed using spectral chunks of size $\Delta z=0.04$ as in \citet{Worseck2011}.
Both observations and simulations show large fluctuations in the effective opacity between different lines of sight. Our numerical results are in good agreement with the observational data while the study by \citet{McQuinn2009} predicts too a steep increase in the effective optical depth with redshift and is therefore unable to match the observed trend.
The different AGN luminosity function used by \citet{McQuinn2009} with respect to our study cannot originate this discrepancy. In fact, the evolution in the number density of \heii\ ionizing photons and of the volume averaged \heii\ fraction are very much consistent between the two theoretical models.
Although a direct comparison between our radiative-transfer calculation and \citet{McQuinn2009} is complicated by the different algorithms and approximations used, we suspect that the discrepancy is mainly due to the different AGN lifetime models. In particular, \citet{McQuinn2009} use a luminosity-dependent AGN lifetime that decreases with increasing AGN luminosity. As a result, the rarest and brightest AGNs at high redshift are short-lived and therefore produce relatively small \heiii\ bubbles. Because the rare, extended regions with low \heii\ opacity at high redshift are typically very highly ionized, they tend to be associated with the proximity regions of the brightest AGNs as we discuss in Section \ref{sec:transparent}.  The distribution of the simulated $\tau_\tr{\heii}^\tr{eff}$ at high redshift is very skew and the value of the mean effective optical depth is mostly influenced by these few transmitted regions in \heii\ spectra. Therefore, a model with a luminosity-independent AGN lifetime and larger \heiii\ bubbles may be able to produce smaller effective optical depths. We will explore in detail the effects of varying the AGN lifetime model in a future work.

\subsection{Statistics}

Additional insight into the probability of observing nearly transparent lines of sight at high redshift is provided in Fig.~\ref{fig:cumulative}, where we show the evolution of the cumulative distribution of the \heii\ effective optical depth in our simulations. Note that this probability is evaluated considering all chunks with $\Delta z =0.04$ (from all simulations) over redshift bins of size $\delta z =0.2$. This helps attenuating short-time fluctuations associated with the variation in the number of ionizing photons emitted by the active sources. Shaded regions correspond to the standard error of the mean between the different realizations.
At high redshift ($z\sim 3.8$), most of the spectra are completely absorbed and only a small fraction of them shows narrow regions with some transmitted flux. Nearly 10 per cent of the spectra have an effective optical depth smaller than 10 and the probability of detecting lines of sight with $\tau_\tr{\heii}^\tr{eff}<2$ is smaller than 1 per mill.
As reionization proceeds, the probability of observing regions with transmitted flux increases.
Approximately $6$ per cent of the sightlines have $\tau_\tr{\heii}^\tr{eff} <3$ at $z\sim 3.4$.
When \heii\ reionization approaches completion ($3.1<z<3.3$, when the \heiii\ filling factor is above $0.95$) the probability of detecting lines of sight with low opacity further increases: nearly $20$ per cent of the sky shows an effective optical depth smaller than $3$ while $\sim 4$ per cent has $\tau_\tr{\heii}^\tr{eff}<2$.
Finally, at $z\sim 3$, as many as 60-70 per cent of the lines of sight are associated with values of the effective optical depth below 3.

\begin{table}
\centering
\begin{tabular}{c c c c c c c}
\hline
 & \multicolumn{2}{c}{$0.995$ CL} & \multicolumn{2}{c}{$0.999$ CL} & \multicolumn{2}{c}{$0.9995$ CL} \\ [0.5ex]
$z$ & $N_\tr{tr}$ & $N_\tr{tot}$ & $N_\tr{tr}$ & $N_\tr{tot}$ & $N_\tr{tr}$ & $N_\tr{tot}$ \\
\hline
$3.4$ & $5$ & $17$ & $7$ & $23$ & $7$ & $23$ \\
$3.6$ & $3$ & $10$ & $4$ & $13$ & $5$ & $17$ \\
$3.8$ & $2$ & $7$  & $3$ & $10$ & $3$ & $10$ \\
$4.0$ & $2$ & $7$  & $2$ & $7$  & $2$ & $7$  \\
\hline
\end{tabular}
\caption{Number $N_\tr{tr}$ of transparent ($\tau_\tr{\heii}^\tr{eff}<3$) and total $N_\tr{tot}$ chunks required at different redshifts to rule out our AGN-driven \heii\ reionization scenario at the $0.995$, $0.999$ and $0.9995$ one-sided Clopper-Pearson confidence levels. It is assumed that the fraction of transparent chunks is $p\sim 0.3$ at all redshifts.}
\label{table:predictions}
\end{table}

For simplicity, we define as `transparent' a chunk characterized by an effective optical depth $\tau_\tr{\heii}^\tr{eff}< 3$. On the other hand, we call opaque a chunk in which the effective optical depth is larger than $100$. In Fig.~\ref{fig:probtau}, we show the probability of measuring a transparent chunk as a function of redshift and directly compare our numerical results with the observational data by \citet{Worseck2014}. Starting from $z>2.9$, we consider three redshift bins of size $\delta z=0.2$ and compute the ratio between the number of transparent and total chunks.
The resulting fraction $p_\tr{obs}$ is shown with a solid symbol.
If we assume that the different chunks are statistically independent (see Section \ref{sec:correlation} for a critical discussion of this working hypothesis), the number of transparent chunks follows a binomial distribution. In this case, $p_\tr{obs}$  is the best estimate for the proportion of transparent chunks in a population, $p$.
In a frequentist sense, confidence intervals for $p$ can be computed using the Clopper-Pearson method \citep[][]{Clopper1934,Gehrels1986}. Error bars in Fig.~\ref{fig:probtau} show the corresponding two-sided $68.3$ and $95.5$ per cent confidence levels for $p$.

\begin{figure}
\centering
\includegraphics[height=0.5\textwidth]{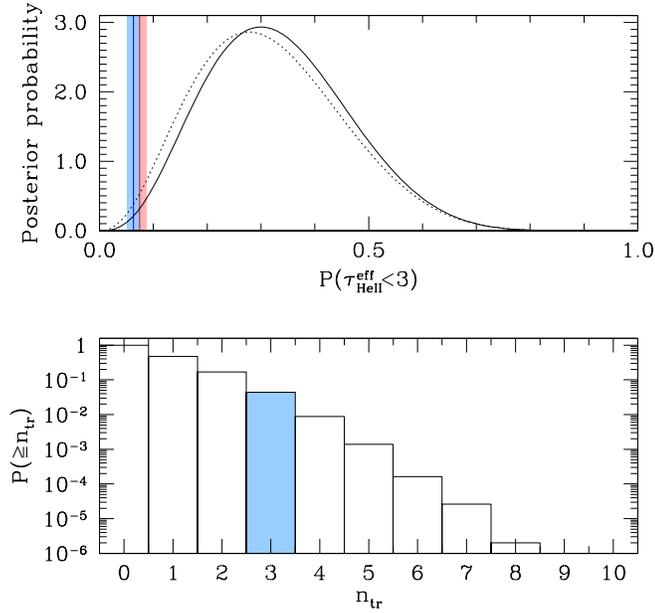}
\caption{Top: posterior probability distribution for the fraction of transparent chunks given the observational data in the redshift bin at $z\sim 3.4$. The solid curve assumes a flat prior, while the dotted curve corresponds to Jeffreys non-informative prior. Vertical lines indicate the results extracted from our numerical simulations without accounting for observability bias (blue) and using the corrective factor discussed in Section \ref{sec:obsbias} (red). Shaded regions denote the $1\sigma$ uncertainty of our model.
Bottom: probability of detecting $n_\tr{tr}$ or more transparent chunks within $10$ chunks selected as in \citet{Worseck2014} in the redshift bin $3.3<z<3.5$.
The value of $n_\tr{tr}=3$ observed with four lines of sight by \citet{Worseck2014} occurs with a probability of $4.4$ per cent in our simulations (shaded area).}
\label{fig:bayesian}
\end{figure}

When reionization is close to completion, simulated data and observations agree well. At higher redshift, the opacity of the IGM in our simulations increases, thus reducing the probability of detecting transparent chunks. This evolution can be noted at $z\sim 3.2$, where observational and simulated data are in excellent agreement suggesting the presence of a higher fraction of \heiii . However, at the highest redshift probed by observations ($z\sim 3.4$), this trend seems to flatten, while in the simulations the probability decreases monotonically with redshift.
Nevertheless, the model predictions lie at the boundary of the 95.67 (two-sided) confidence level for the data
(which corresponds to $\sim 2$ standard deviations, $\sigma$, in a Gaussian distribution). Therefore, the model can only be rejected with relatively low statistical significance.
Note, however, that it is impossible to determine exact confidence level for a discrete distribution like the binomial one \citep[][]{Neyman1935} and several studies have shown that the Clopper-Pearson method is too conservative with coverage probabilities that are often higher than their nominal values \citep[e.g.][]{Vollset1993,Agresti1998,Cameron2011}.
This implies that the deviation between the data and the model might be more significant than what we found using the Clopper-Pearson method. Alternatively, we perform a Bayesian analysis in the bin at $z\sim 3.4$ and evaluate the posterior distribution of $p$ given the observational data (see the top panel in Fig.~\ref{fig:bayesian}). We consider two different priors: a flat one and Jeffreys non-informative prior for the binomial distribution.
When tested in the frequentist sense, the credibility intervals associated with these posterior distributions provide better coverage probabilities than the Clopper-Pearson method \citep{Cameron2011}. We find that the posterior probability that $p$ is lower than what is predicted by our models is $0.4$ and $0.8$ per cent for the flat and Jeffreys priors, respectively. In a Gaussian distribution, this corresponds to $\sim 2.7$ and $\sim 2.4\sigma$ (one-sided) deviations.

\begin{table*}
\centering
\begin{tabular}{c@{\extracolsep{0.5cm}} c c@{\extracolsep{0.5cm}} c c@{\extracolsep{0.5cm}} c c@{\extracolsep{0.5cm}} c c@{\extracolsep{0.5cm}} c}
\hline
 & & \multicolumn{2}{c}{Clopper-Pearson} & \multicolumn{2}{c}{Flat prior} & \multicolumn{2}{c}{Jeffreys prior} & \multicolumn{2}{c}{Monte Carlo} \\ [0.5ex]
$N_\tr{tr}$ & $N_\tr{tot}$ & CL & CL$_\tr{bias}$ & CL & CL$_\tr{bias}$ & CL & CL$_\tr{bias}$ & CL & CL$_\tr{bias}$ \\
\hline
$3$ & $10$ & $97.840$ & $96.650$ & $99.633$ & $99.334$ & $99.236$ & $98.715$ & $95.601$ & $93.521$ \\
$6$ & $20$ & $99.886$ & $99.733$ & $99.979$ & $99.942$ & $99.958$ & $99.892$ & $99.490$ & $98.909$ \\
$9$ & $30$ & $99.993$ & $99.976$ & $99.999$ & $99.995$ & $99.997$ & $99.990$ & $99.932$ & $99.797$ \\
\hline
\end{tabular}
\caption{One-sided frequentist confidence levels and Bayesian credibility levels (per cent) at which our model at $z=3.4$ is ruled out after observing $N_\tr{tot}$ chunks with $\Delta z =0.04$ out of which $N_\tr{tr}$ are transparent. Results obtained including the corrective factor for the observability bias are indicated with the subscript `bias'.}
\label{table:confidence}
\end{table*}

In Table~\ref{table:predictions}, we report the number of transparent and total chunks necessary to rule out our AGN-driven \heii\ reionization scenario at the $0.995$, $0.999$ and $0.9995$ (one-sided) Clopper-Pearson confidence levels.
For each redshift bin, we select an integer value $N_\tr{tr}$ of transparent chunks and evaluate the corresponding number of total chunks $N_\tr{tot}$ for which the fraction $N_\tr{tr}/N_\tr{tot}$ is closest to the value of $0.3$ which has been measured at $z\sim 3.4$. We repeat this calculation increasing the number of transparent chunks until the results from our simulations are not anymore compatible with the observational data at the selected confidence level. We find that, at the $0.999$ confidence level (corresponding to roughly $3$ Gaussian $\sigma$) the detection of four additional transparent chunks would be necessary to rule out our reionization model. Observational measurements at $z>3.4$ would require a smaller amount of lines of sight, but the number of unabsorbed quasars suitable for \heii\ studies drastically decreases.
Note that our estimate assumes that $p\sim 0.3$ at all redshifts. If future observations would detect a lower $p$ (as evidenced by our simulations), the values reported in Table~\ref{table:predictions} are underestimated. In this case, additional transparent chunks from \heii\ quasars would be required to reliably confirm the importance of sources of hard UV photons other than AGNs in the \heii\ reionization.

\subsubsection{Observability bias} \label{sec:obsbias}

In the previous analysis we have considered the probability of detecting a transparent chunk from a set of randomly drawn lines of sight in our simulation boxes and compared it to observations. In reality, however, the selection of observed lines of sight is not random: any intervening, optically thick \hi\ system may extinguish the quasar continuum at or below the \heii\ edge, rendering the \heii\ opacity measurements impossible.
Indeed, because of the accumulated absorption by \hi\ systems in the spectra of $z>3$ quasars, only a few per cent of them - the so-called \heii\ quasars - are suitable targets for \heii\ studies \citep[see e.g.][]{Worseckand2011}.
The vast majority of intervening systems are at much lower redshift than the regions probed by the \heii\ spectra. However, the \heii\ quasar selection technique also rejects targets with \hi\ absorbers at $z>3$ that are thick enough to extinguish the quasar far-UV flux \citep{Worseckand2011}. For instance, considering a spectral chunk at $z=3.4$, all lines of sight containing a Lyman limit system (LLS) with $\tau_\tr{\hi}^\tr{eff}>5$ at the same redshift will be extinguished at the \heii\ edge and therefore not included in the observed sample. This introduces an \emph{observability bias} that is redshift dependent and that is not included in the randomly selected sample of lines of sight in our simulations. Clearly, the effect of the observability bias is to remove \heii\ opaque lines of sight from the sample and therefore boost the fraction of transparent chunks with respect to a randomly selected sample.
How important is this effect? Unfortunately, current large cosmological simulations cannot directly resolve the LLSs, and this is especially true for our simulations that are aimed for volume rather than resolution. Therefore, we cannot directly reproduce the observability bias with our simulations. However, we can estimate the fraction $f_{\mathrm{LLS}}$ of lines of sight in our computational boxes that should contain a LLS with $\tau_\tr{\hi}>5$ [or $\log (N_\tr{\hi}/\tr{cm}^{-2})>17.9$] using the observed differential number density of LLSs as a function of redshift \citep{Prochaska2010,Songaila2010}. We then rank our $n_{\mathrm{tot}}$ simulated lines of sight by their total $N_\tr{\hi}$ in decreasing order and remove from our sample the first $f_{\mathrm{LLS}}\times n_{\mathrm{tot}}$, with the very plausible assumption that the lines of sight with the highest column density would be associated with LLSs at higher resolution. Once we make this correction, we obtain a fraction of simulated transparent chunks at $z\sim 3.4$ that is a factor $1.2$ higher than for randomly selected lines of sight, further relieving the tension between the model and the observations.
In fact, the predictions of our simulations now lie within the 93.3 per cent (two-sided) Clopper-Pearson confidence level (corresponding to $\sim 1.8\sigma$ in the Gaussian case). On the other hand, the Bayesian posterior probability grows to $0.7$ and $1.3$ per cent (i.e. $\sim 2.5$ and $\sim 2.2\sigma$ for the Gaussian statistics). Note that this factor is not included in the model lines presented in Fig.~\ref{fig:probtau}.

\subsubsection{Accounting for statistical correlations}\label{sec:correlation}

The $10$ chunks observed at $z\sim 3.4$ are extracted  from four quasar spectra and the transparent ones come from only two lines of sight \citep{Worseck2014}. This suggests that the assumption of statistical independence between the chunks is likely to be violated so that the binomial distribution does not apply. In order to quantify the importance of this effect, we use a Monte Carlo method to resample four random lines of sight from our simulations and extract $10$ chunks out of them with the same pattern (i.e. number of considered chunks per observed sightline) as in \citet{Worseck2014}. The resulting cumulative probability distribution is shown in the bottom panel of Fig.~\ref{fig:bayesian}.
While there is a considerable probability of detecting a number of transparent chunks equal to or higher than one or two ($47.6$ and $17.0$ per cent, respectively), the chance of matching the observational data ($n_{\rm tr}\geq 3$) is of $4.4$ per cent. This corresponds to the probability above $1.7\sigma$ in a Gaussian distribution, showing that correlations among the data make the deviation of our model less statistically significant than what is found using the binomial approximation\footnote{We have checked that considering one chunk per sightline in our simulations perfectly reproduces the binomial predictions showing
that the finite size of the simulation box does not introduce any detectable correlations.}.
Replicating this analysis while accounting for the observability bias, the probability of matching the observational data increases to $6.5$ per cent or, equivalently, $1.5\sigma$ in a Gaussian distribution. In Table~\ref{table:confidence}, we report the one-sided confidence levels and credibility levels at which our model of \heii\ reionization is ruled out according to the different statistical methods considered in this work. We select integer multiples of the number of transparent and total chunks obtained from observations at $z\sim 3.4$ and evaluate the confidence levels between these data and the results from our simulations for the Clopper-Pearson analysis, the Bayesian analysis (both with flat and Jeffreys priors) and the Monte Carlo analysis. In the last case, the chunks in the lines of sight are selected duplicating the observed pattern, as previously discussed. For each method we include the results obtained considering the corrective factor for the observability bias (indicated by the subscript `bias').
Based on this study, we conclude that the tension between the observational data and the AGN-driven scenario of \heii\ reionization is mild and additional data are required to rule out the model with a convincing statistical significance.

\begin{figure}
\centering
\includegraphics[height=0.5\textwidth]{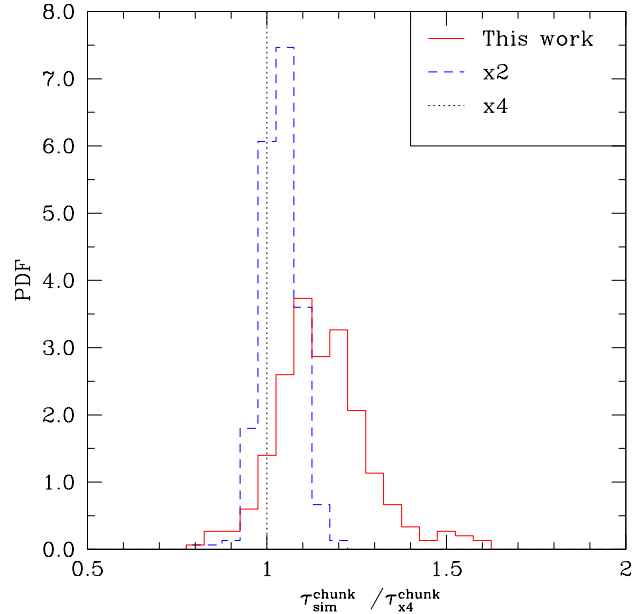}
\caption{Quantification of resolution effects on the measurement of $\tau_\tr{\heii}^\tr{eff}$ (measured in bins of $\Delta z=0.04$ at $z=3.4$) in numerical simulations. We consider three AMR simulations in which the same realization of the IGM is evolved using resolution elements with different sizes for the base grid (see main text for details). We consider the same resolution used for our study of helium reionization (red solid) and two other resolution elements whose linear size is a factor of 2 (blue dashed) and 4 (black dotted) smaller. Plotted is the PDF of the ratio between $\tau_\tr{\heii}^\tr{eff}$ evaluated for the same chunk at a given resolution and at the maximum resolution.}
\label{fig:resolution}
\end{figure}

\subsubsection{Numerical resolution effects}\label{sec:resolution}

\begin{table}
\centering
\begin{tabular}{@{\extracolsep{0.5cm}}l c @{\extracolsep{0.5cm}}c c @{\extracolsep{0.5cm}}c}
\hline
Method & CL & $\sigma$ & CL$_\tr{bias}$ & $\sigma_\tr{bias}$\\
\hline
Clopper-Pearson & $91.14$ & $1.35$ & $87.63$ & $1.16$ \\
Flat prior      & $97.44$ & $1.95$ & $95.91$ & $1.74$ \\
Jeffreys prior  & $95.84$ & $1.73$ & $93.75$ & $1.53$ \\ 
Monte Carlo     & $86.67$ & $1.11$ & $82.25$ & $0.93$ \\
\hline
\end{tabular}
\caption{One-sided frequentist confidence levels and Bayesian credibility levels (per cent and equivalent number of Gaussian standard deviations) at which our model at $z=3.4$ is ruled out by the observational data in \citet{Worseck2014} after accounting for finite-resolution effects in our simulations. Results obtained including the corrective factor for the observability bias are indicated with the subscript `bias'.}
\label{table:resolution}
\end{table}

The \heii\ Ly$\alpha$ transition has a large cross-section (relative to hydrogen) and transmission through the IGM is primarily associated with low-density gas \citep[see e.g][]{Rauch1998}. Due to their finite spatial resolution, numerical simulations tend to overestimate the density in these regions and to overpredict the corresponding \heii\ effective optical depth  \citep[see e.g.][for a discussion of this effect in smoothed particle hydrodynamic simulations]{Theuns1998}. In order to account for systematic errors, we study how $\tau_\tr{\heii}^\tr{eff}$ at $z=3.4$ varies with the resolution of the base grid of our AMR simulations. We start from the base mesh given in Table \ref{table:sims} and we decrease the linear size of the resolution element by a factor of $2$ and $4$. We process these three boxes with the {\sevensize RADAMESH} code and extract spectra of the \heii\ Ly$\alpha$ forest in redshift bins of $\Delta z=0.04$ probing exactly the same density field at different resolutions.  As a source of ionizing radiation we use a uniform background whose amplitude is fixed to reproduce the mean transmissivity\footnote{Note that a uniform radiation field produces less optically thick regions than a set of discrete sources.} found in the simulations with radiative transfer discussed above.
A convergence study is provided in Fig. \ref{fig:resolution}, where we show the probability distribution function of the ratio between the effective optical depths evaluated, for the same chunk, in a given simulation and in the simulation with the highest resolution. This test suggests that the simulations with radiative transfer adopted in this work might overestimate the \heii\ optical depth by about $15$ per cent on average. Therefore resolution effects should further alleviate the tension between our simulations and the observational data at $z\sim 3.4$. We approximately quantify their impact by convolving the PDF of $\tau_\tr{\heii}^\tr{eff}$ determined in Section \ref{sec:optical_depth} with the correction given in Fig. \ref{fig:resolution}. The corresponding frequentist confidence levels and the Bayesian credibility levels at which our model is ruled out are reported in Table \ref{table:resolution} for the different statistical methods considered in this work, where the subscript `bias' indicates the values obtained considering the corrective factor for the observability bias. 
We conclude that the discrepancy between the observational data by \citet{Worseck2014} and our simulations is only a $1\sigma$ to $2\sigma$ effect.

\begin{figure}
\centering
\includegraphics[height=0.5\textwidth]{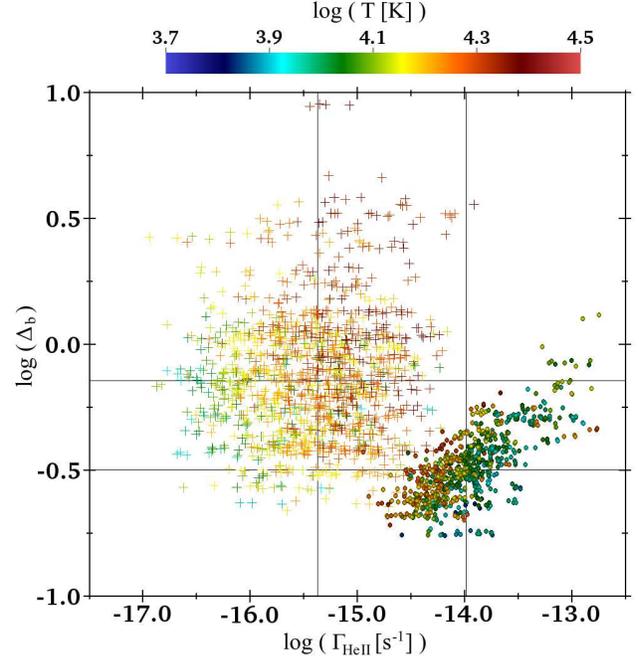}
\caption{Distribution of the helium photoionization rate $\Gamma_\tr{\heii}$ and the baryonic overdensity $\Delta_\tr{b}$ for gas elements associated with transparent (circles) and opaque (crosses) chunks of size $\Delta z=0.04$ in the redshift interval $3.3<z<3.5$.
Points are colour-coded based on the logarithm of the gas temperature as indicated by the colour bar.
Solid lines indicate the median values of the distributions.}
\label{fig:proj}
\end{figure}

\subsection{The nature of transmitted chunks at $z=3.4$}\label{sec:transparent}

The opacity at $304\,\tr{\AA}$ of a gas element is directly related to the fraction of \heii\ in the cell which, in turn, can be linked to the interplay between the density (governing the rate of recombinations), the local flux of ionizing photons (responsible for the rate of photoionizations) and the temperature of the gas.
In Fig.~\ref{fig:proj}, we compare the properties of transmitted and opaque chunks extracted from our mock spectra in the redshift interval $3.3<z<3.5$.
We associate to each transparent chunk a representative cell that corresponds to the maximum transmitted flux in the chunk (circles), while a random cell within the chunk is associated to each opaque region of the spectrum (crosses). The scatter plot presents the relationship between the photoionization rate $\Gamma_\tr{\heii}$ and the baryonic overdensity $\Delta_\tr{b}$ and is colour-coded with respect to the temperature $T$ of the gas element. Cells associated with transparent chunks are predominantly located in mean or underdense regions (mainly because of the selection criterion) and show a large spread in the values of the temperature. The photoionization rate is systematically larger than what is observed for opaque chunks. These completely absorbed regions extend to higher density and are described by an average temperature that is to some extent higher with respect to the transmitted case.

In order to evaluate the origin of the transparent regions in \heii\ spectra, we can associate to each representative cell the source that contributes the most to the local $\Gamma_\tr{\heii}$.
The average distance between this source and the representative cell is generally smaller ($\sim 13.6\,\tr{\mh}$ comoving) for transparent chunks with respect to the value observed for opaque regions ($\sim 17.2 \,\tr{\mh}$) or for a random distribution of points in the computational box ($\sim 16.3 \,\tr{\mh}$). In roughly half of the transparent ($57$ per cent) and opaque ($49$ per cent) chunks, the fraction of the total ionizing flux supplied by this single AGN is higher than $90$ per cent. Although one source generally provides the main contribution, cells associated with transparent chunks often receive photons from several different AGNs located nearby, thus being characterized by a large total photoionization rate ($\Gamma_\tr{\heii}>10^{-15}\,\tr{s}^{-1}$ for all cells associated with transparent chunks). On the other hand, opaque chunks are generally affected by one single source and only a small fraction of them receive a significant number of ionizing photons. On the grounds of this evidence, we can infer that regions of transmitted flux in \heii\ Ly$\alpha$ spectra at high redshift can be originated by two different effects. The first possibility arises when the line of sight crosses the proximity region (i.e. the zone in which the ionizing flux due to the dominant source is larger than the contribution due to all other AGNs) of a single luminous AGN. In this case, the large photoionization rate within the ionization front produces transparent chunks that can be ascribed to the so-called `transverse proximity effect'. A second, less common, possibility occurs when the transmitting region is close to several, fainter, AGNs. In this case the combined photoionization rate produced by the sources is strong enough to generate a transparent chunk.

\section[]{CONCLUSIONS}
We perform hydrodynamic simulations of the epoch of \heii\ reionization adopting the same technique presented in Paper~I: the output of an AMR run is post-processed using a modified version of the radiative-transfer code {\sevensize RADAMESH} \citep{Cantalupo2011}.
We use the quasar luminosity function observed in the redshift range $3.5<z<5.2$ to calibrate the magnitude of the sources in our simulated volumes, producing a realistic population of AGNs that turn on at $z_\tr{AGN}=5$. We consider different models for the redshift evolution of the emissivity of the sources and perform several runs to account for the different ionization history associated with the position and luminosity of the AGNs. We extract simulated spectra of the \heii\ Ly$\alpha$ forest from our computational volumes and compare our results with the most recent observational data available to date \citep{Worseck2014}.

In agreement with our previous theoretical work and with observations of the intergalactic \heii\ absorption in \ti{HST/COS} spectra, we find that the observed quasar population can completely ionize most of the \he\ in the IGM by $z\sim 3$. All computational volumes show that the \heii\ fraction drops below $0.5$ per cent in the redshift interval $2.9 \leq z \leq 3.1$, with a \heiii\ filling factor larger than $99.9$ per cent for $z<2.8$.
The completion of the reionization seems to be unaffected by the exact start of the radiative-transfer calculations as long as a sufficiently extended reionization process is considered.
However, the increasing gas density and the decline of the AGN abundance towards high redshifts cause a slower evolution of the IGM properties during the early stages of \heii\ reionization with respect to the results of Paper~I.

The effective optical depth of the \heii\ Ly$\alpha$ forest extracted from simulated spectra matches the observational data by \citet{Worseck2014} at all redshifts. Additional sources of hard UV photons beyond AGNs are not required to explain current data. However, moderate tension (at about the Gaussian $2\sigma$ level) is present at the highest redshift which has been currently probed by observations ($z\sim 3.4$). Accounting for resolution effects in our simulations further alleviates the tension with the observational data (the significance of the discrepancy drops between $1\sigma$ and $2\sigma$ depending on the statistical method adopted).

For $z>3$, our model predicts a large variability between different lines of sight generated by the ongoing reionization process as found by observations. At these redshifts, $\tau_\tr{\heii}^\tr{eff}$ evaluated in chunks of $\Delta z=0.04$ has a very broad and skew distribution.
The median effective optical depth increases at high redshifts more slowly than previous numerical simulations \citep[e.g.][]{McQuinn2009}, accounting for small, but rare, values of $\tau_\tr{\heii}^{\tr{eff}}$ up to $z\sim 4$. The fluctuations in the effective optical depth predicted by our model correlate with large variations in the temperature of the medium that persist after the completion of the reionization process.

Transparent lines of sight at high redshifts are generated by the joint influence of the density, the temperature and, mostly, the local photoionization rate. Regions of high transmissivity in \heii\ spectra are usually underdense and cross the proximity regions of luminous sources capitalizing on the transverse proximity effect generated by bright AGNs. In other cases, the transmitted region is generated by the contribution of several, less luminous, sources that provide the required number of ionizing photons. During the final phases of \he\ reionization, the temperature of these transparent regions is generally slightly lower than what is measured for regions of high opacity. This reflects the fact that transparent regions have been ionized earlier and are characterized by a \heii\ fraction of $f_\tr{\heii}\sim 10^{-3}$ or lower, while opaque regions are still experiencing ionization.

The moderate tension between the simulated and observed $\tau_\tr{\heii}^{\tr{eff}}$ may be due to the observability bias, a systematic lack of UV photons in the models (associated with the poorly constrained shape of the AGN luminosity function, lifetime and spectral energy distribution at high redshift), or, simply, a statistical fluctuation in the observational data. 
In order to rule out our model with high significance, the sample size at $z\sim 3.4$ needs to be at least two or three times bigger than it is now. If future data will confirm the current measurements, there will be sufficient evidence for considering additional hard UV radiation and, maybe, other sources beyond AGNs.

A key challenge in a complete characterization of the reionization process remains: constraining the evolution of the \heii\ effective optical depth at redshift $z\geq 3.5$. When a sufficient number of lines of sight towards \heii\ quasars will become available in the future, tighter constraints on the timing, topology and the sources of \heii\ reionization will be placed from the joint analysis of simulated and observed transparent lines of sight.

\section*{ACKNOWLEDGEMENTS}
We thank an anonymous referee for helpful comments which improved the quality of this work.
MC thanks Gabor Worseck for providing the observational data before publication and David Syphers for his constructive comments and suggestions on an earlier version of the paper.
Numerical simulations were run at the Leibniz-Rechenzentrum centre in Munich.
This work is carried out within the Collaborative Research Centre 956, sub-project C4, funded by the Deutsche Forschungsgemeinschaft (DFG).
SC acknowledges support from the NSF AST-1010004 and STScI HST GO-13013.04-A grants.


\appendix

\section[]{SOFT UV BACKGROUND} \label{app:uvbackground}

\begin{figure}
\centering
\includegraphics[width=0.5\textwidth]{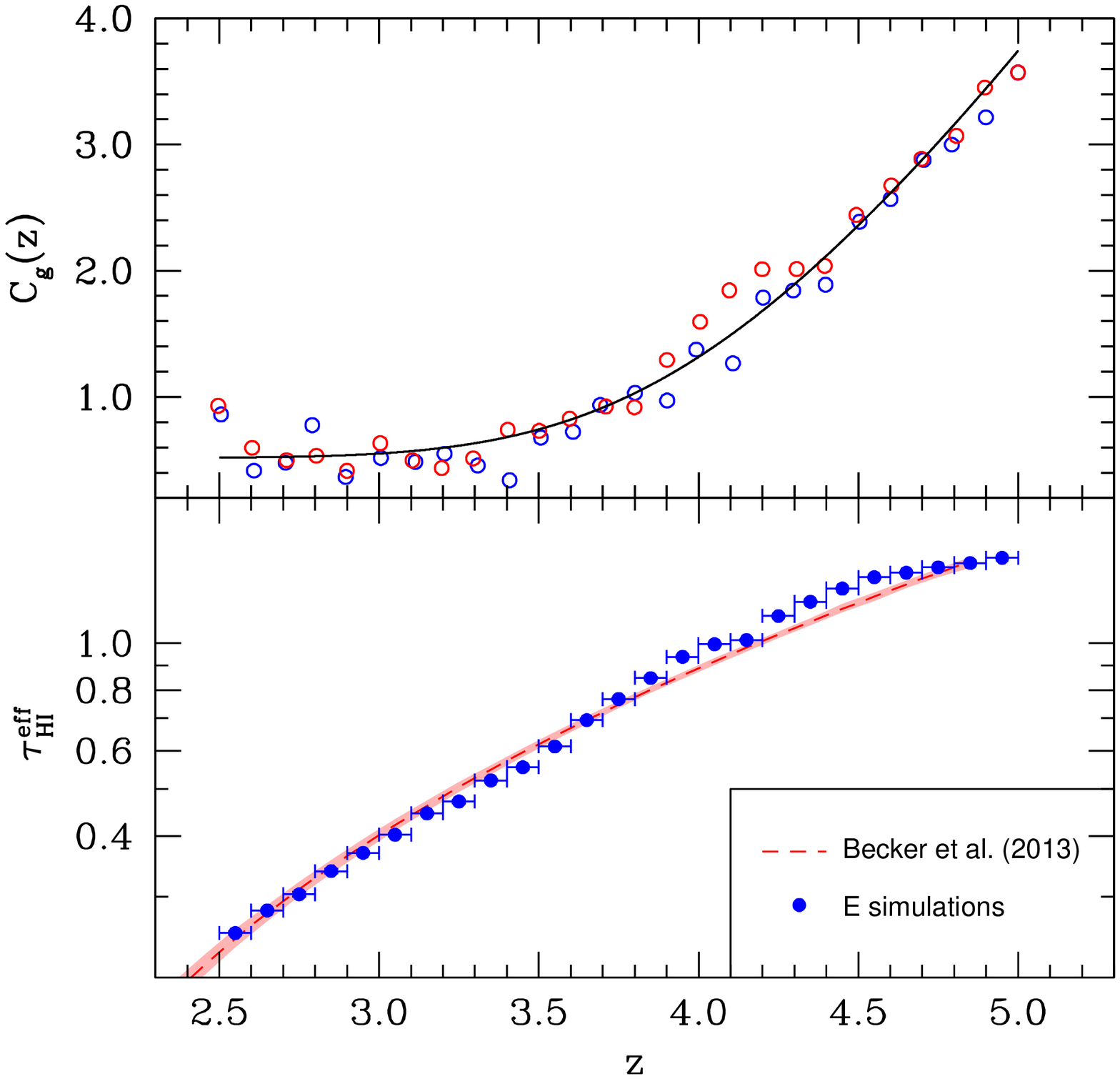}
\caption{Top: redshift evolution of the multiplicative correction factor $C_\tr{g}(z)$. Circles represent the value of $C_\tr{g}(z)$ that must be applied to two test simulations (PLE in blue and PDE in red) in order to match observations of the \hi\ effective optical depth. The solid line represents the fit given in equation (\ref{eq:correction}).
Bottom: redshift evolution of the \hi\ effective optical depth in our simulations evaluated using bins of $\Delta z=0.1$. The analytic fit to observational data (red dashed line) by \citet{Becker2013} and its $1\sigma$ uncertainty (shaded region) are also reported for comparison.
}
\label{fig:meantauH}
\end{figure}

Our simulations use the soft UV background computed by \citet{Haardt2012} as a template. Assuming that the temperature of the IGM is $20\,000$ K, this model reproduces the observed evolution of the effective optical depth in the \hi\ Ly$\alpha$ line \citep{Haardt2012}.
However, most of the gas in our simulations never becomes so hot. For instance, the average gas temperature at mean density is of approximately $6500$ K at $z=5$ and does not exceed $17\,000$ K even when it reaches its peak value during \heii\ reionization at $z\sim 3.4$. Since the \hi\ opacity strongly depends on the temperature of the absorbing gas, we need to use a different UV background to match the observed evolution of the IGM transmissivity.
One possibility would be to leave the photoionization rate unchanged and increase the photoheating rate. However, this would require the presence of much harder sources of radiation which are difficult to reconcile with the assumption that only young stars contribute at early times. We therefore prefer to rescale the predictions by \citet{Haardt2012} with a redshift-dependent multiplicative factor, $C_\tr{g}(z)$, which is determined requiring that the \hi\ transmissivity of the Ly$\alpha$ forest reproduces the most recent observational measurements by \citet{Becker2013}.
To estimate the amplitude of the correction factor, we run two radiative-transfer simulations of \heii\ reionization using the soft UV background computed by \citet{Haardt2012} combined with our two source models for AGNs (PLE or PDE). We then use the fluctuating Gunn-Peterson approximation \citep{Croft1998} to scale the IGM opacity in the simulations in order to match the data by \citet{Becker2013}.
We find that the expression
\begin{equation} \label{eq:correction}
C_\tr{g}(z)=0.26\left[\tanh(z-3.8)+1\right](z-2.4)^{2}+0.52
\end{equation}
provides a good analytical approximation to the numerical results (see the top panel in Fig.~\ref{fig:meantauH}).

As a cross check, the bottom panel in Fig.~\ref{fig:meantauH} shows that the \hi\ effective optical depth extracted from our simulations matches the observational data points basically at all times. Note, however, that the opacity extracted from the mock spectra is very sensitive to random fluctuations in the number and luminosity of the ionizing sources.
For instance, the slight excess of opacity seen at $z\sim 4.4$ must be ascribed to a temporary decrease in the number of ionizing photons emitted within the computational volumes per unit time.


\bsp
\label{lastpage}
\end{document}